# Axion Insulator State in a Ferromagnet/Topological Insulator/Antiferromagnet Heterostructure

Y. S. Hou and R. Q. Wu

Department of Physics and Astronomy, University of California, Irvine, CA 92697-4575, USA

## Abstract

We propose to use ferromagnetic insulator $MnBi_2Se_4$/$Bi_2Se_3$/antiferromagnetic insulator $Mn_2Bi_2Se_5$ heterostructures for the realization of the axion insulator state. Importantly, the axion insulator state in such heterostructures only depends on the magnetization of the ferromagnetic insulator and hence can be observed in a wide range of external magnetic field. Using density functional calculations and model Hamiltonian simulations, we find that the top and bottom surfaces have opposite half-quantum Hall conductance, $\sigma_{xy}^{t} = e^2/2h$ and $\sigma_{xy}^{b} = -e^2/2h$, with a sizable global spin gap of 5.1 meV opened for the topological surface states of $Bi_2Se_3$. Our work provides a new strategy for the search of axion insulators by using van der Waals antiferromagnetic insulators along with three-dimensional topological insulators.

Email: wur@uci.edu

Three-dimensional (3D) topological insulators (TIs) with metallic topological surface states (TSSs) protected by the time-reversal symmetry [1, 2] have become prototypes for the realization of many emergent physical properties such as the quantum anomalous Hall effect [3-5], quantized magneto-optical effect [5-9] and Majorana fermion state [10]. One of the most striking but underexplored phenomena is the topological magnetoelectric (TME) effect in the magnetized 3D TIs that is evoked by the topological term $\theta$ in the axion electrodynamics [5, 11-20]. Simply put, this peculiar effect manifests in the so-called axion insulators as that an external magnetic (electric) field induces an electric (magnetic) polarization [5]. Such mutual controls between magnetic and electric polarizations make axion insulators promising candidates for spintronics and quantum information operations [21, 22]. Experimentally, the realization of axion insulators require a special magnetic configuration that the magnetization over the surfaces of 3D TIs points either all inwards or all outwards so that their TSSs are globally gapped [5, 23]. In practice, this is achieved by doping the two surfaces of 3D TIs with different magnetic elements, which produces different coercive fields in two sides [5, 11, 23-26]. As one sweeps the external magnetic field, the axion insulator state (AIS) forms as the magnetizations of two surfaces align antiparallelly. However, since the coercive fields of doped TIs are not easy to control and dopants may redistribute with time in samples, the AIS produced in this way is difficult to control and sustain. It is hence desirable to put forward new strategies to design stable and controllable axion insulators.

In this Letter, we propose using ferromagnetic (FM) insulator/3D-TI/antiferromagnetic (AFM) insulator (FM/TI/AFM) heterostructures to make more stable AIS. As we only operate on the magnetization of the FM part, the AIS in such heterostructures may appear at zero external magnetic field. Based on density functional theory (DFT) calculations and low-energy effective four-band model simulations, we demonstrate that AIS indeed shows up in the FM $MnBi_2Se_4$ (MBS) septuple layer (SL)/3D TI $Bi_2Se_3$ (BS)/AFM $Mn_2Bi_2Se_5$ ($M_2BS$) nonuple layer (NL) heterostructures. Our work thus introduces a new strategy for searching axion insulators.

DFT calculations are carried out by using the Vienna *Ab Initio* Simulation Package at level of the generalized gradient approximation [27-30]. We use the projector-augmented wave pseudopotentials to describe the core-valence interaction [31, 32] and set the energy cutoff of 500 eV for the plane-wave expansion [30]. Ten quintuple layers (QLs) of BS are used for building up the MBS/BS/MBS, $M_2BS/BS/M_2BS$ and $MBS/BS/M_2BS$ slabs and the vacuum space between the adjacent slabs is 12 Å. The weak interaction across $MBS/M_2BS$ and BS layers is described by the nonlocal van der Waals (vdW) functional (optB86b-vdW) [33, 34]. To consider the correlation effect among Mn 3*d* electrons, we employ the LSDA+U method with the on-site Coulomb interaction U=6.0 eV and the exchange interaction J=1.0 eV [35, 36]. Spin-orbit coupling is included in the calculations of all band structures.

To study the topological properties of various heterostructures, we utilize a low-energy effective four-band model. This model consists of the TSSs ($H_{\text{surf}}$) of 3D TI BS, the exchange field ($H_{\text{Zeeman}}$) from magnetic MBS and $M_2BS$ and the interfacial potential ($H_{\text{Interface}}$). With the basis set of $\{|t,\uparrow\rangle,|t,\downarrow\rangle,|b,\uparrow\rangle,|b,\downarrow\rangle\}$, the low-energy effective four-band model is [3, 37-39]

$$H(k_x,k_y) = H_{\text{surf}}(k_x,k_y) + H_{\text{Zeeman}}(k_x,k_y) + H_{\text{Interface}}(k_x,k_y)$$

$$= A(k_x^2+k_y^2) + \begin{bmatrix} 0 & iv_Fk_- & 0 & 0 \\ -iv_Fk_+ & 0 & 0 & 0 \\ 0 & 0 & 0 & -iv_Fk_- \\ 0 & 0 & iv_Fk_+ & 0 \end{bmatrix} + \begin{bmatrix} \Delta_t & 0 & 0 & 0 \\ 0 & -\Delta_t & 0 & 0 \\ 0 & 0 & \Delta_b & 0 \\ 0 & 0 & 0 & -\Delta_b \end{bmatrix} + \begin{bmatrix} V & 0 & 0 & 0 \\ 0 & V & 0 & 0 \\ 0 & 0 & -V & 0 \\ 0 & 0 & 0 & -V \end{bmatrix} \quad (1)$$

Here, *t* (*b*) denotes the top (bottom) surface state and ↑(↓) represent the spin up (down) states; $v_F$ is the Fermi velocity; $k_x$, $k_y$ and $k_\pm = k_x \pm ik_y$ are wave vectors; $\Delta_t$ and $\Delta_b$ are the exchange fields experienced by the top and bottom TSSs, respectively; 2*V* is the inversion asymmetric interface potential. In MBS/BS/MBS and $M_2BS/BS/M_2BS$, the inversion symmetry is kept so $\Delta_t = \Delta_b = \Delta$ and V=0. However, the inversion symmetry is broken in $MBS/BS/M_2BS$. Note that the hybridization between the top and bottom TSSs is not included in this model as the BS film is very thick (10 QLs). Berry curvatures and Chern numbers ($C_N$) are calculated based on the formulas in Ref. [40, 41].

We first demonstrate the advantage of the FM/TI/AFM heterostructure for the realization of AIS over the conventional FM/TI/FM heterostructure [5, 11]. In the latter case, the top and bottom surfaces need to have antiparallelly aligned magnetizations with an external magnetic field $\mu_0 H$ in the range of $\mu_0 H_{2c} < |\mu_0 H| < \mu_0 H_{1c}$ (Fig. 1a). Obviously, it is rather challenging to observe the AIS in such heterostructure when the coercive field difference $\mu_0|H_{1c} - H_{2c}|$ is small [23, 42]. Since layered AFM insulators play the same role as FM insulators in magnetizing TSSs [36, 43], we propose to realize the AIS in the FM/TI/AFM heterostructure. When the FM insulator and the first-layer magnetic ions of the layered AFM insulator have opposite magnetizations (insets in Figs. 1b and 1c), the TSS of the TI film is globally gaped. Furthermore, because the layered AFM insulator has no net magnetization, its magnetic structure is not affected when sweeping the external magnetic field in a wide range. Consequently, the magnetic alignment between the FM and AFM parts can be easily reset by applying an external magnetic field $\mu_0 H > \mu_0 H_{1c}$ ($H_{1c}$ is the coercive field of the FM insulator) (Fig. 1b) and the AIS in the FM/TI/AFM heterostructure may appear at zero external magnetic field and in a broad range $|\mu_0 H| < \mu_0 H_{1c}$.

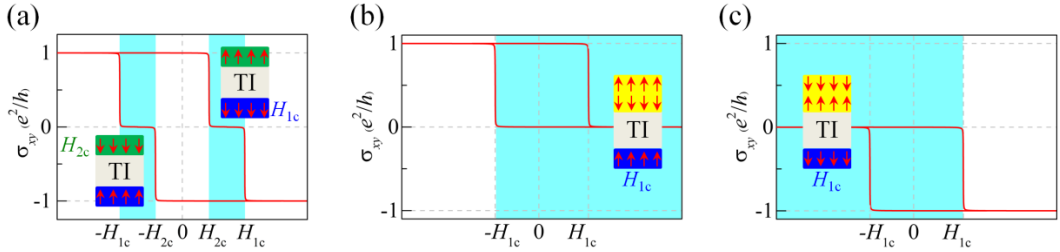

*Figure 1 (color online) (a) External field dependence of the Hall conductivity $\sigma_{xy}$ of a FM/TI/FM heterostructure. (b) and (c) External field dependence of the Hall conductivity $\sigma_{xy}$ of a FM/TI/AFM heterostructure. The cyan regions indicate regions of the external magnetic field that give rise to the axion insulate state; the insets show the magnetization configurations of AIS. Red arrows represent the magnetic moments of FM and AFM insulators.*

There are two factors that are crucial for the realization of the AIS in the FM/TI/AFM heterostructure. (1) The AFM insulators should open a spin gap for the TSSs of 3D TIs but not damage their transport properties. It is nontrivial as previous studies showed that most layered AFM insulators such as MnSe [36, 43] hybridize too strongly with TIs and the Dirac points are shifted far away from the Fermi level. (2) The remaining hybridization between the top and bottom TSSs needs to be eliminated to obtain the quantized TME effect [11, 24]. This requires that the 3D TI films are thick enough. For BS, noticeable hybridization between the two surfaces starts as the TI films is thinner than 6 QLs [44-47]. In experiments, it is difficult to define the thickness of TI well, due to the existence of rotation twins [48, 49] but we recommend to use thick TI films for the observation of AIS. From the band structures (Part I in Supplementary material (SM) [50]), we find that the hybridization between the top and bottom TSSs is negligible across ten QLs of BS as we adopt in the present studies.

For the best match with BS, we pick MBS SL and $M_2BS$ NL as the FM and layered AFM insulators, respectively. A recent experiment shown that FM MBS SL (Fig. 2a) could be synthesized by depositing a MnSe bilayer onto the BS's topmost QL [51]. DFT calculations confirmed that MBS SL is an energetically stable FM insulator [51]. Based on these findings, Our DFT calculations indicate that $M_2BS$ NL has a lower energy by 55.6 meV per atom than the MBS SL/MnSe bilayer heterostructure (Part II in SM) and we may hence assume that $M_2BS$ NL can be synthesized as well by depositing a MnSe bilayer onto MBS SL. After considering several different collinear spin orders and the typical noncollinear $120^o$-AFM order [52] (Part III in SM), we find that $M_2BS$ NL has a layered AFM order (Fig. 3a). Moreover, this AFM $M_2BS$ NL is an insulator with a gap of 0.6 eV (Part III in SM). Lastly, our DFT calculations show both MBS SL and $M_2BS$ NL have an out-of-plane magnetic anisotropy (Part IV in SM).

Fig. 2b shows the band structure of MBS/BS/MBS. In consistent with previous studies [51], the TSSs of BS are magnetized by the FM MBS SL and a large spin gap of 52.2 meV is opened. Such a large gap is due to the large extension of the TSSs of BS into the FM MBS SL as they share common lattice structures and most chemical compositions (Fig. 2a) [52]. The vdW gap at the MBS/BS interface is $d_1$=2.66 Å, which is markedly smaller than the vdW gap (about 3.20 Å) in $CrI_3$/BS/$CrI_3$ [53]. The two-fold degenerate bands near the Fermi level around the Γ point (left inset in Fig. 2b) confirm that there is no noticeable hybridization between the top and bottom TSSs. By fitting the DFT band structures to the low-energy effective four-band model Eq. (1) (Part V in SM), we obtain a Chern number $C_{N,1}=-1$ (Fig. 2c). This nontrivial $C_{N,1}=-1$ is also obtained by calculations with the Wannier90 package [54] (Part V in SM). This indicates the applicability of the low-energy effective four-band model in the present work.

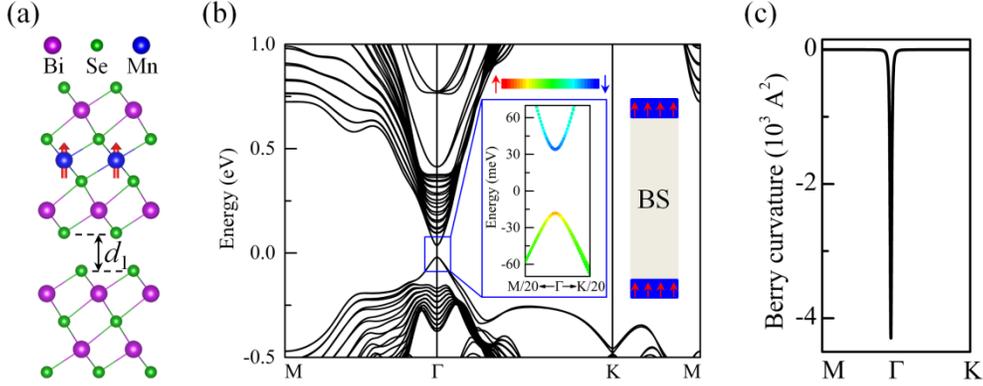

*Figure 2 (color online) (a) Schematic atomic structure at the MBS/BS interface of MBS/BS/MBS. Magnetic moments of Mn in the FM MBS SL are represented by the red arrows. The vdW gap at the interface is shown by $d_1$. (b) Band structure of MBS/BS/MBS. Left inset shows the spin-projected bands near the Fermi level with spin weights being indicated by the color bar. Right inset sketches the magnetitic configuration. (c) Berry curvature of the occupied bands calculated from the effective four-band model.*

Now we examine if the layered AFM $M_2BS$ NL may also magnetize the TSSs of BS. As shown by the band structure of $M_2BS/BS/M_2BS$ in Fig. 3b, a spin gap of 6.4 meV opens. Importantly, the Fermi level locates in the gap of the magnetized TSSs so there is no detrimental effect on the transport properties of BS in $M_2BS/BS/M_2BS$. This is completely different from the BS/MnSe heterostructure where the magnetized TSSs are shifted away from the Fermi level [36, 43]. Again, both the four-band model with fitting parameters and Wannier90 package give a Chern number of $C_{N,2}=-1$ for $M_2BS/BS/M_2BS$ (Part VI in SM). The vdW gap in $M_2BS/BS/M_2BS$ is also small ($d_2$=2.62 Å), but the spin gap of the TSSs is no as large as that in MBS/BS/MBS because of the mutual cancellation between opposite contributions from the two Mn layers. If we set M$_2$BS NL in the FM state, the spin gap of TSSs of M$_2$BS/BS/M$_2$BS increases to 51.3 meV (Part VII in SM).

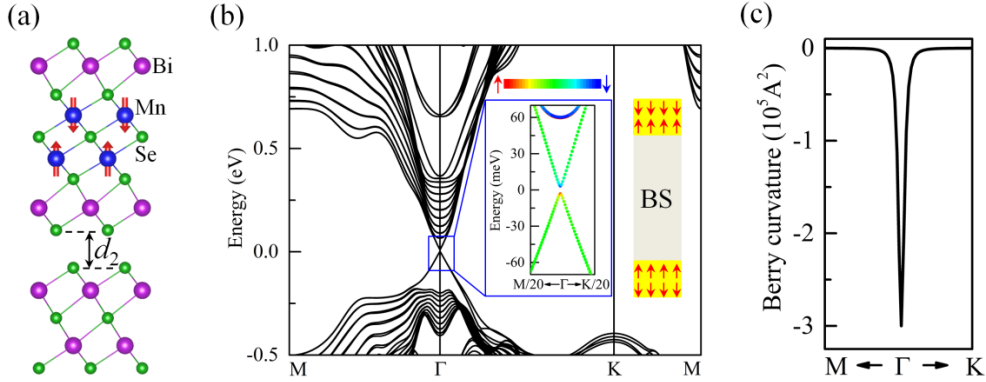

*Figure 3 (color online) (a) Schematic atomic structure at the M$_2$BS/BS interface of M$_2$BS/BS/M$_2$BS. Magnetic moments of Mn in the layered AFM M$_2$BS NL are represented by the red arrows. The vdW gap at the interface is shown by d$_2$. (b) Band structure of M$_2$BS/BS/M$_2$BS. Left inset shows the spin-projected bands near the Fermi level and spin weights are indicated by the color bar. Right inset sketches the magnetitic configuration. (c) Berry curvature of the occupied bands calculated from the effective four-band model.*

Finally, we assemble the FM-MBS/BS/AFM-M$_2$BS heterostructure and explore the possibility of achieving the AIS. We set the magnetization of the FM MBS SL in the upward direction while the magnetization of the first Mn layer in the AFM M$_2$BS NL in the downward direction, as shown in the right inset in Fig. 4a. The band structure clearly suggests a global spin gap of 5.1 meV across the Fermi level, showing the appropriate magnetization of TSSs (Fig. 4a). By projecting bands to the top and bottom QLs, we observe that the magnetized TSSs from the top and bottom surfaces are well separated (Fig. 4b). Fig. 4b and the left inset of Fig. 4a suggest the spin components of TSSs from the top and bottom surfaces around the $\Gamma$ point. Interestingly, the four branches of TSSs at the top and bottom surfaces sense opposite exchange fields, i.e., $\Delta_t \Delta_b < 0$.

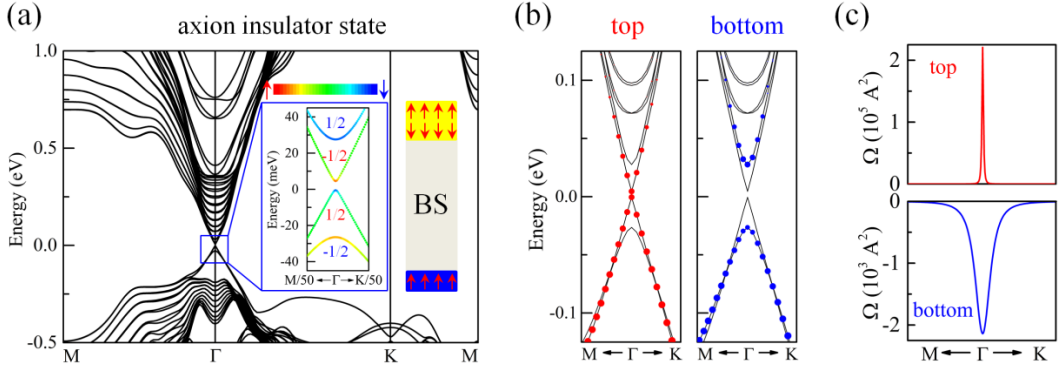

Figure 4 (color online) (a) Band structure of MBS/TI/M$_2$BS in the AIS. Left inset shows the spin-projected four bands near the Fermi level and their Chern numbers. Right inset sketches the magnetitic configuration. Color bar indicates the spin-projected weights. (b) Top-QL-BS and bottom-QL-BS projected band structures. (c) Berry curvatures of the magnetized TSSs below $E_F$ for the top (up panel) and bottom (bottom panel) surfaces.

By fitting the DFT band structure using Eq. (1), we obtain that (1) the top and bottom exchange fields are $\Delta_t = 2.6$ meV and $\Delta_b = -26.8$ meV, respectively; (2) the asymmetric interface potential $2V$ is 1.0 meV. Furthermore, the occupied TSSs from the top and bottom surfaces have opposite Berry curvatures, as shown in Fig. 4c. Integrating the Berry curvatures of the occupied bands in the Brillouin zone gives that the Chern numbers of the top and bottom surfaces are $C_N^t = 1/2$ and $C_N^b = -1/2$, respectively. In other words, the top and bottom surfaces have opposite half-quantum Hall conductances $\sigma_{xy}^t = e^2/2h$ and $\sigma_{xy}^b = -e^2/2h$, respectively. This is the direct evidence of the AIS in MBS/TI/M$_2$BS. The vanishing global Hall conductance $\sigma_{xy} = \sigma_{xy}^t + \sigma_{xy}^b = e^2/2h - e^2/2h = 0$ is further verified by calculations using the Wannier90 package (Part VIII in SM). The small asymmetric interface potential $2V$ due to the difference between FM-MBS and AFM-M$_2$BS appears not to affect the topological properties.

Obviously, MBS/BS/M$_2$BS can also be easily transformed to the Chern insulator state when the magnetization of MBS SL is reversed from upward to downward (Part IX in SM). As the spin orientation of the FM layer is reset, the Berry curvature of the occupied TSSs at the MBS/BS surface flips. Integrating the Berry curvatures of the occupied bands in the Brillouin zone gives rise to the global Chern number $C_N = 1$ (Part IX in SM), a signature of the quantum anomalous Hall effect. Obviously, it is very convenient to switch MBS/BS/M$_2$BS between the Chern insulator and axion insulator states.

To summarize, we demonstrated that MBS/BS/M$_2$BS heterostructure may manifest stable AIS using DFT calculations and model simulations. As MBS and similar systems have been successfully grown in experiments [51, 55], we believe that the fabrication of MBS/BS/M$_2$BS heterostructure is feasible. Our work significantly expands the range of searching axion insulators, from conventional FM/TI/FM heterostructures to FM/TI/AFM heterostructures.

Work was supported by DOE-BES (Grant No. DE-FG02-05ER46237). Density functional theory calculations were performed on parallel computers at NERSC supercomputer centers.

# Supplementary Material of "Axion Insulator States in a Ferromagnet/Topological Insulator/Antiferromagnet Heterostructure"

Y. S. Hou and R. Q. Wu

Department of Physics and Astronomy, University of California, Irvine, CA 92697-4575, USA

## Part I. Band structure of ten quintuple layers (QLs) of $Bi_2Se_3$

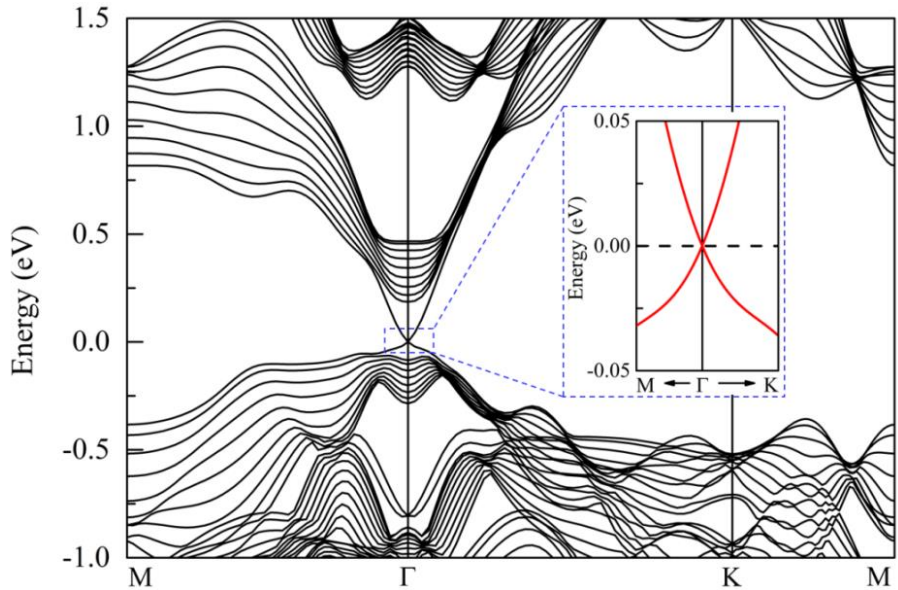

Figure S1 (color online) DFT calculated band structure of ten QLs of $Bi_2Se_3$. The metallic topological Dirac cone near the Fermi level is zoomed in the inset. The gap is zero.

# Part II. Comparison between $Mn_2Bi_2Se_5$ nonuple layer and $MnBi_2Se_4$ Septuple layer (SL)/MnSe bilayer (BL) heterostructure

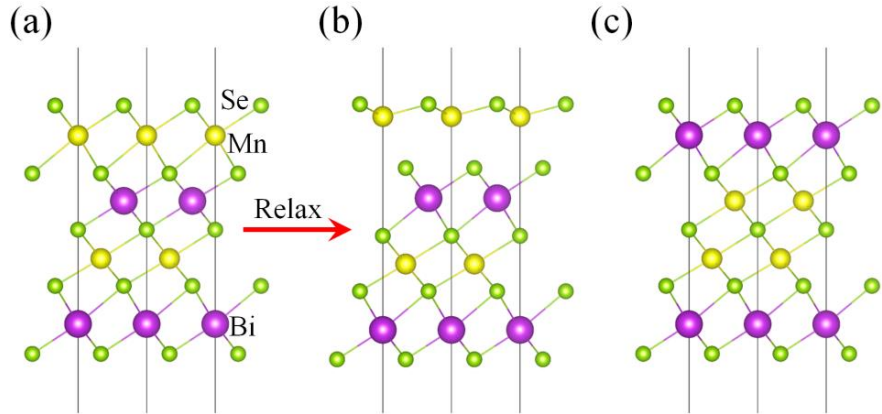

Figure S2 (color online) Side views of (a) the initial and (b) the relaxed atomic structure of the $MnBi_2Se_4$ SL/MnSe BL heterostructure. (c) Side view of the atomic structure of the relaxed $Mn_2Bi_2Se_5$ NL.

## Part III. Magnetic grounds of Mn$_2$Bi$_2$Se$_5$ NL

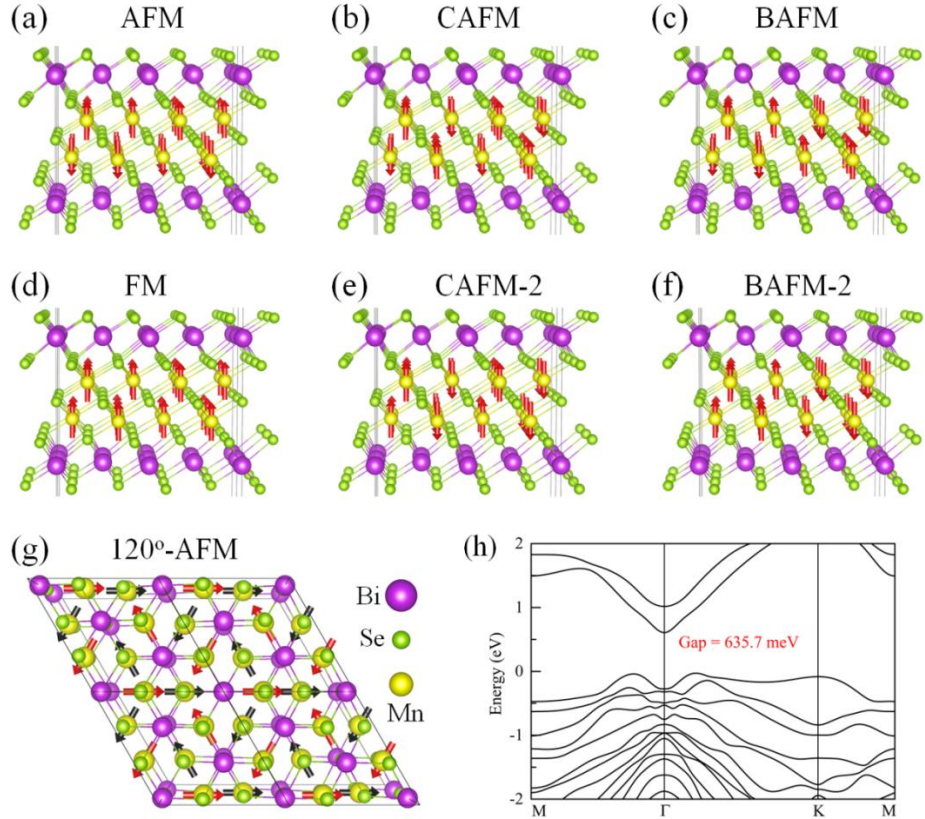

Figure S3 (color online) (a)-(f) Six different collinear spin orders and (g) the noncollinear 120o-AFM spin order of Mn$_2$Bi$_2$Se$_5$ NL. In (a)-(g), arrows represent magnetic moments. In (g), magnetic moments of the upper and lower layers of Mn ions are represented by the red and black arrows, respectively. (h) The DFT calculated band structure of the AFM spin order as shown in (a) of Mn$_2$Bi$_2$Se$_5$ NL.

Table S1 Energies of the seven spin orders as shown in the Figure S3. The ground state spin order AFM (Figure S3a) is set as the reference.

| Spin order | AFM | CAMF | BAFM | FM | CAFM-2 | BAFM-2 | 120°-AFM |
|---|---|---|---|---|---|---|---|
| Energy (meV/Mn) | 0 | 3.8 | 2.2 | 1.6 | 5.8 | 2.5 | 4.5 |

**Part IV. Magnetic easy axes of MnBi$_2$Se$_4$ SL and Mn$_2$Bi$_2$Se$_5$ NL**

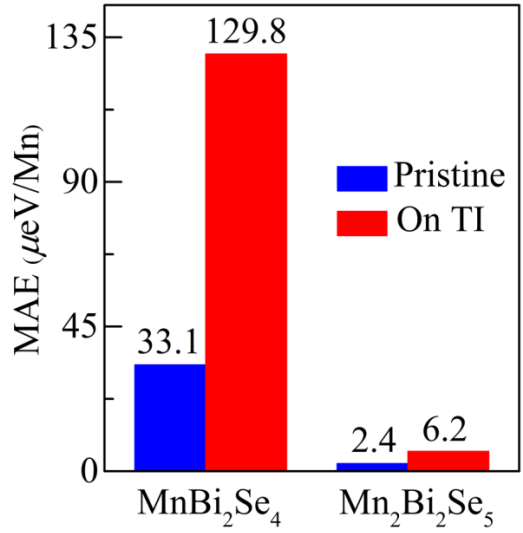

Figure S4 (color online) Magnetocrystalline anisotropic energy (MAE) of MnBi$_2$Se$_4$ SL and Mn$_2$Bi$_2$Se$_5$ NL. The MAE is defined as E$_{MAE}$=E(*x*)-E(*z*), where E(*x*) and E(*z*) are the energies with magnetic moments of Mn atoms being in-plane and out-of-plane. From this figure, we can get that both MnBi$_2$Se$_4$ SL and Mn$_2$Bi$_2$Se$_5$ NL have an out-of-plane magnetic easy axis. It is worth noting that their out-of-plane MAEs are strongly enhanced when MnBi$_2$Se$_4$ SL and Mn$_2$Bi$_2$Se$_5$ NL are contacted with TI BS.

**Part V. Topological properties of MnBi$_2$Se$_4$/Bi$_2$Se$_3$/MnBi$_2$Se$_4$ heterostructure**

We use the effective four-band model as shown in the Eq. (1) in the main text to fit the DFT calculated band structure and Berry curvature of MnBi$_2$Se$_4$/Bi$_2$Se$_3$/MnBi$_2$Se$_4$ heterostructure. Because the inversion symmetry is preserved in the this heterostructure, the exchange field parameters $\Delta_t$ ($\Delta_b$) are same and set to be $\Delta$. Note that this argument is also applied to the inversion symmetric Mn$_2$Bi$_2$Se$_5$/Bi$_2$Se$_3$/Mn$_2$Bi$_2$Se$_5$ heterostructure. For MnBi$_2$Se$_4$/Bi$_2$Se$_3$/MnBi$_2$Se$_4$ heterostructure, the fitting parameters are A=10.10 eVÅ$^2$, $v_F$ =1.71 eVÅ, and $\Delta$=0.026 eV.

In this work, we use Wannier90 package to numerically calculate the Chern number $C_N$ based on the formula $C_N = \frac{1}{2\pi} \frac{S_{FBZ}}{N} \sum_{i=1}^{N} \Omega_i$. Here $S_{FBZ}$, $N$ and $\Omega$ are the area of first Brillouin zone (FBZ), the number of sampled points in FBZ and Wannier functions calculated Berry curvature. Note that the points are evenly distributed in FBZ in Eq. (A1). If the points are not even distributed, their weight should be taken into account. For MnBi$_2$Se$_4$/Bi$_2$Se$_3$/MnBi$_2$Se$_4$, 10000 points are evenly distributed in the FBZ (Figure S6) and we get $C_{N,1}$=-1. Especially, we note the magnitude of the Berry curvatures calculated based on the effective four-band model fitting and by the Wannier90 package are almost same (see Figure S5 (b) and Figure S6 (b); Figure S7 (b) and Figure S8 (b)). This indicates that our model fitting is of super high quality.

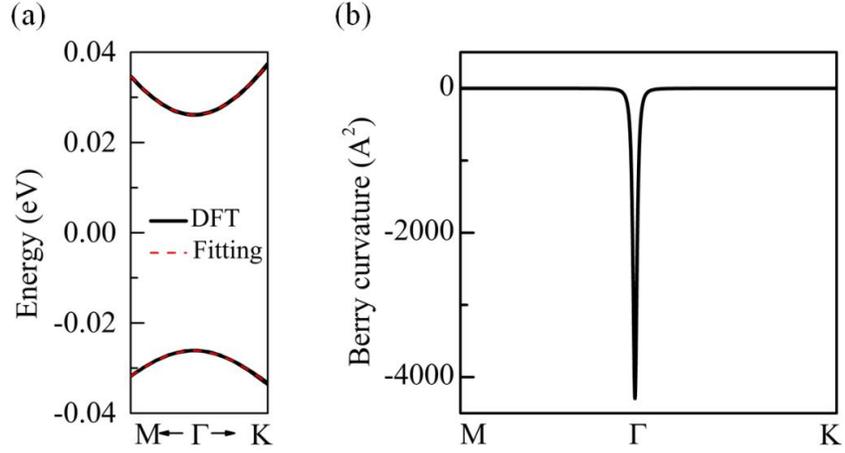

Figure S5 (color online) (a) DFT calculated (black lines) and fitted (red dashed lines) band structures of $MnBi_2Se_4/Bi_2Se_3/MnBi_2Se_4$ heterostructure. (b) Berry curvature of the occupied bands of $MnBi_2Se_4/Bi_2Se_3/MnBi_2Se_4$ heterostructure is calculated based on the effective four-band model as shown in the Eq. (1) in the main text.

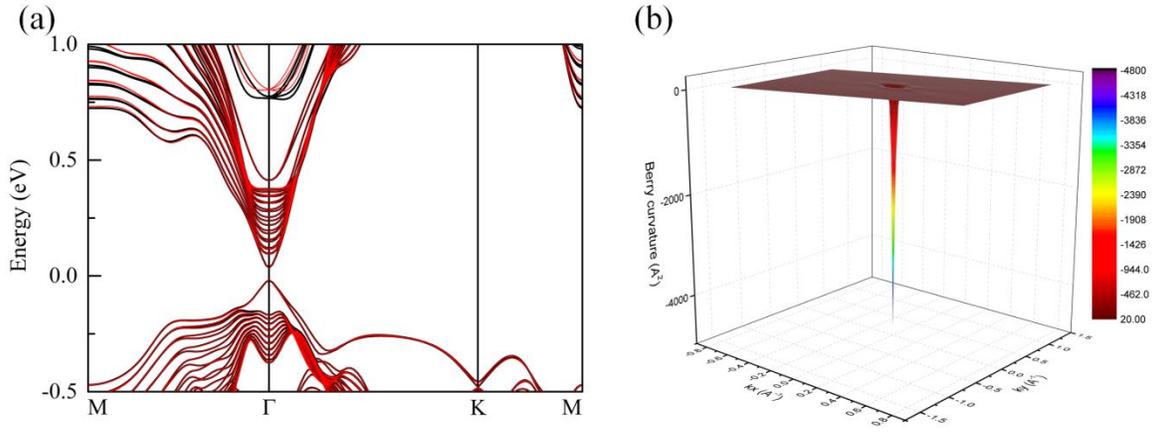

Figure S6 (color online) (a) DFT calculated (black lines) and Wannier90 interpolated (red lines) band structures of $MnBi_2Se_4/Bi_2Se_3/MnBi_2Se_4$ heterostructure. (b) Berry curvature in the FBZ is calculated by Wannier90 package.

# Part VI. Topological properties of $Mn_2Bi_2Se_5/Bi_2Se_3/Mn_2Bi_2Se_5$ heterostructure

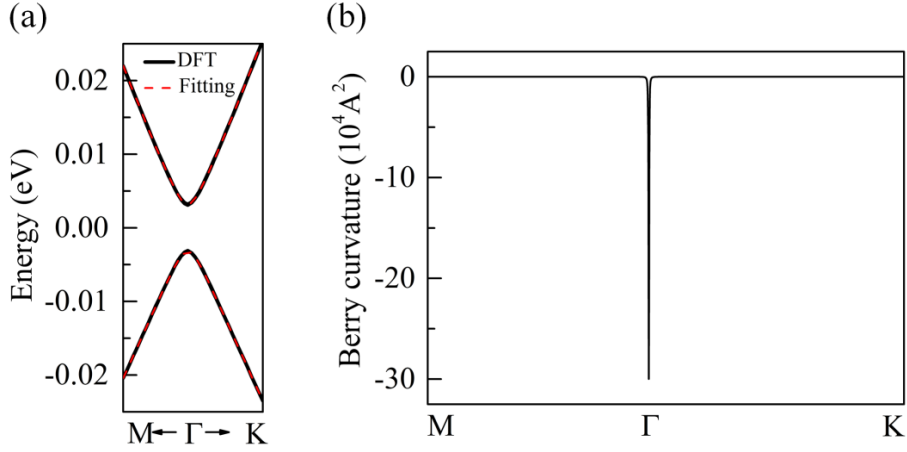

Figure S7 (color online) (a) DFT calculated (black lines) and fitted (red dashed lines) band structures of $Mn_2Bi_2Se_5/Bi_2Se_3/Mn_2Bi_2Se_5$ heterostructure. (b) Berry curvature of the occupied bands of $MnBi_2Se_4/Bi_2Se_3/MnBi_2Se_4$ heterostructure is calculated based on the effective four-band model as shown in the Eq. (1) in the main text. The fitting parameters are A=5.33 eVÅ$^2$, $v_F$=1.74 eVÅ, and Δ=0.0032 eV.

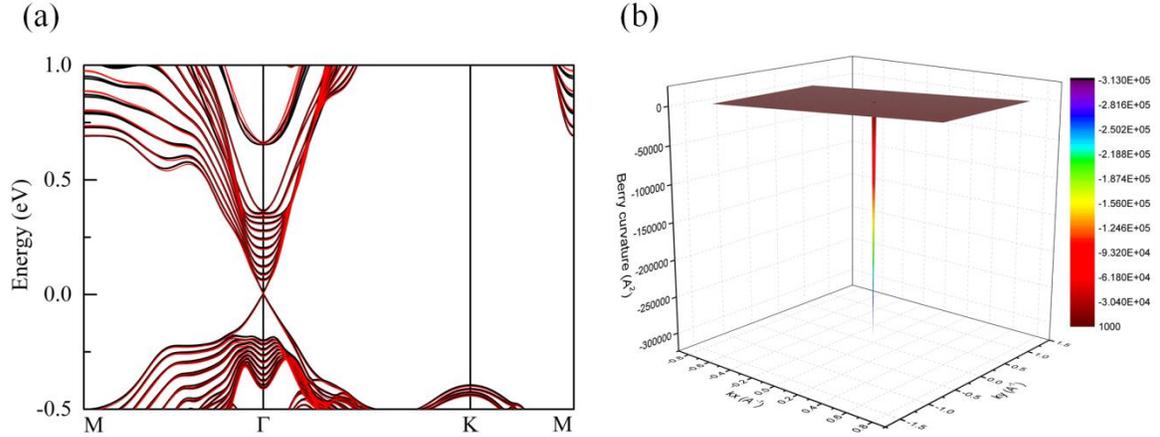

Figure S8 (color online) (a) DFT calculated (black lines) and Wannier90 interpolated (red lines) band structures of $Mn_2Bi_2Se_5/Bi_2Se_3/Mn_2Bi_2Se_5$ heterostructure. (b) Berry curvature in the FBZ is calculated by Wannier90 package. Here 10000 points are sampled around the Γ point and 10000 points are sampled elsewhere and we get $C_{N,2}$=-1.

# Part VII. Band structure of $Mn_2Bi_2Se_5/Bi_2Se_3/Mn_2Bi_2Se_5$ heterostructure with ferromagnetic order in $Mn_2Bi_2Se_5$ NL

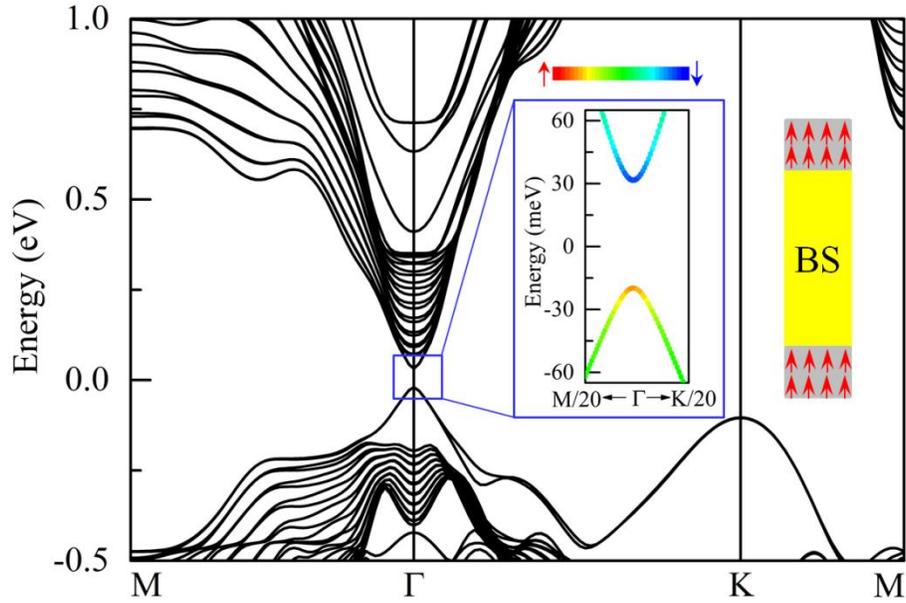

Figure S9 (color online) DFT calculated band structures of $Mn_2Bi_2Se_5$/BS/$Mn_2Bi_2Se_5$ heterostructures with ferromagnetic order in $Mn_2Bi_2Se_5$ NL. The left inset shows the spin-projected four bands near the Fermi level whereas the right inset sketch the magnetization configurations. Spin-projected weights are indicated by the color bar.

# Part VIII. Wannier90 package calculated Berry curvature of the axion insulate state of MnBi$_2$Se$_4$/Bi$_2$Se$_3$/Mn$_2$Bi$_2$Se$_5$ heterostructure

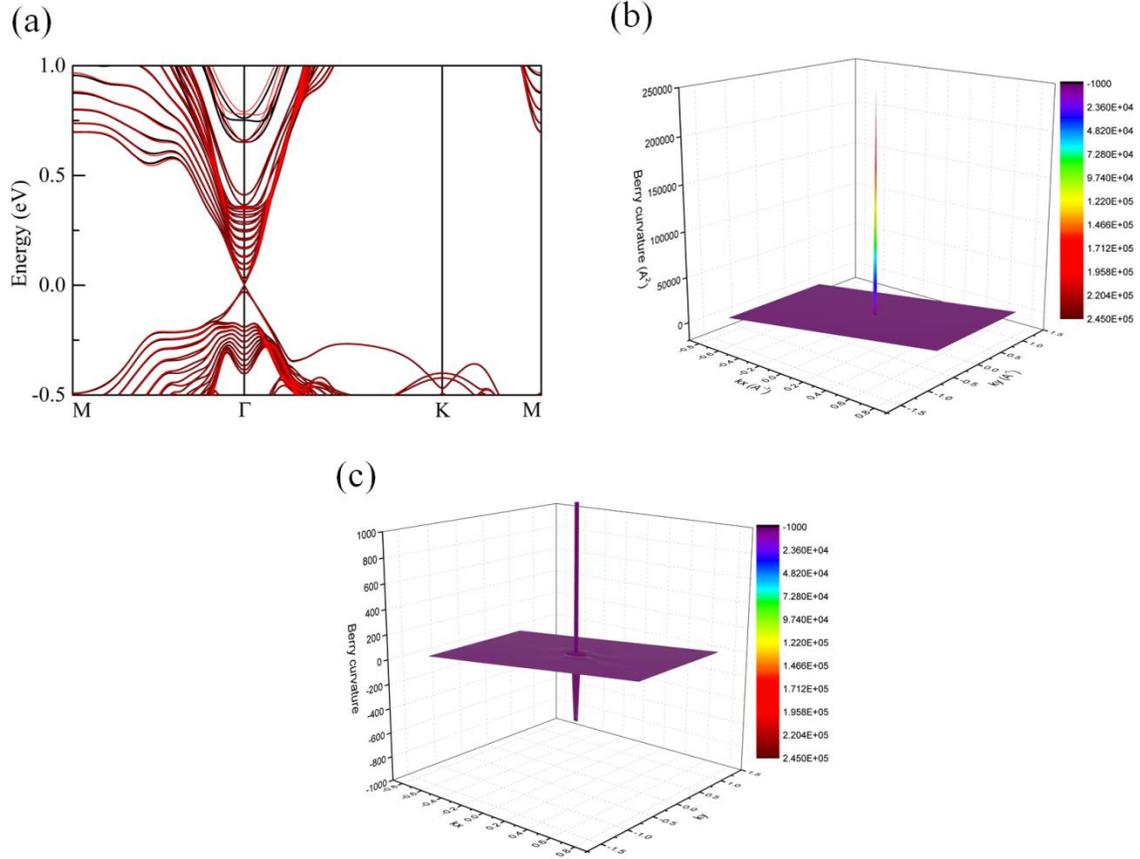

Figure S10 (color online) (a) DFT calculated (black lines) and Wannier90 interpolated (red lines) band structures of the axion insulator state of MnBi$_2$Se$_4$/Bi$_2$Se$_3$/Mn$_2$Bi$_2$Se$_5$ heterostructure. (b) Berry curvature in the FBZ is calculated by Wannier90 package. (c) Berry curvature in in the range from -1000 to 1000 A$^2$. We can see that there is non-negligible negative Berry curvature except the extremely large positive Berry curvature as shown in (b). Here 10000 points are sampled around the Γ point and 10000 points are sampled elsewhere and we get $C_N$=0. This is consistent with the zero Hall conductance as obtained in the main text.

# Part IX. Chern insulator state in the MnBi$_2$Se$_4$/Bi$_2$Se$_3$/Mn$_2$Bi$_2$Se$_5$ heterostructure

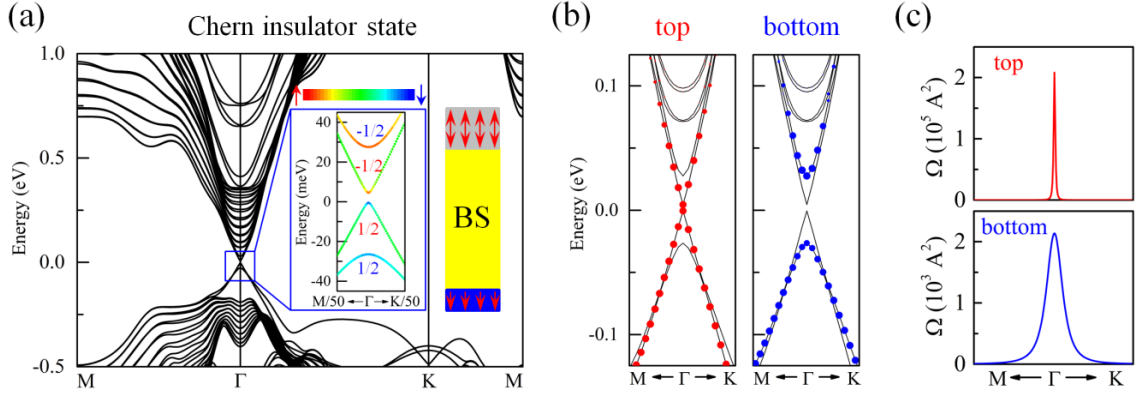

Figure S11 (color online) (a) Band structure of Chern insulator state in MnBi$_2$Se$_4$/TI/Mn$_2$Bi$_2$Se$_5$ heterostructure. Left inset shows the spin-projected four bands near the Fermi level and their Chern numbers whereas right inset sketches the magnetization configurations. Color bar indicates the spin-projected weights. (b) Top-QL-BS and bottom-QL-BS projected band structures are shown by the red dots in the left panel and the blue dots in the right panel, respectively. (c) Berry curvatures ($\Omega$) of the occupied magnetized TSSs of the top (up panel) and bottom (bottom panel) surfaces.

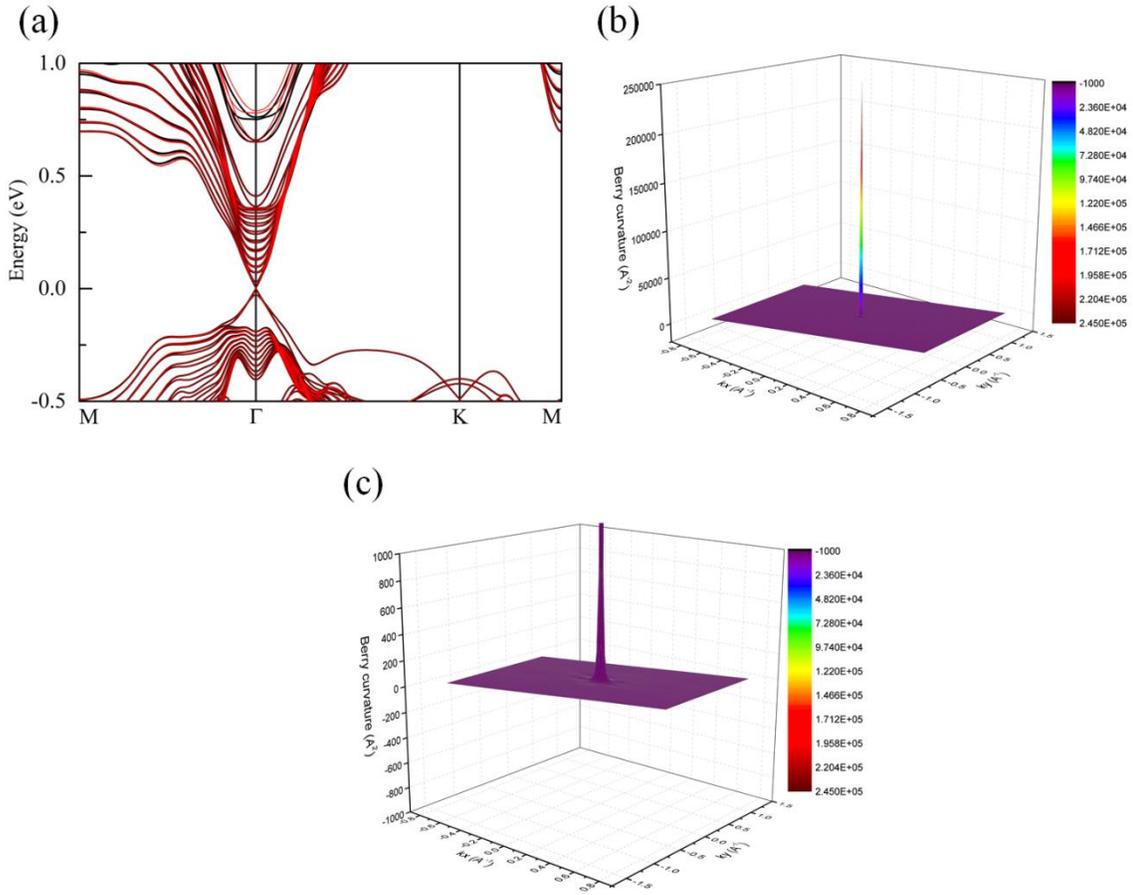

Figure S12 (color online) (a) DFT calculated (black lines) and Wannier90 interpolated (red lines) band structures of the Chern insulator state of MnBi$_2$Se$_4$/Bi$_2$Se$_3$/Mn$_2$Bi$_2$Se$_5$ heterostructure. (b) Berry curvature in the FBZ is calculated by Wannier90 package. (c) Berry curvature in in the range from -1000 to 1000 A$^2$. We can see that there is no non-negligible negative Berry curvature except the extremely large positive Berry curvature as shown in (b). Here 10000 points are sampled around the Γ point and 10000 points are sampled elsewhere and we get $C_N$=1. This is consistent with the zero Hall conductance as obtained in the main text.